\documentclass[12pt]{article}
\usepackage{epsfig,amsfonts,amssymb}
\begin{document}

\voffset -0.7 true cm
\hoffset 1.1 true cm
\topmargin 0.0in
\evensidemargin 0.0in
\oddsidemargin 0.0in
\textheight 8.6in
\textwidth 7.1in
\parskip 10 pt

\def\ket#1{\vert #1 \rangle}
\def\bra#1{\langle #1 \vert}

\newcommand{\be}{\begin{equation}}
\newcommand{\ee}{\end{equation}}
\newcommand{\bea}{\begin{eqnarray}}
\newcommand{\eea}{\end{eqnarray}}
\newcommand{\beas}{\begin{eqnarray*}}
\newcommand{\eeas}{\end{eqnarray*}}

\begin{titlepage}
\begin{flushright}
{\small CU-TP-1053} \\
{\small hep-th/0203083}
\end{flushright}

\begin{center}

\vspace{1cm}

{\Large \bf de Sitter entropy from conformal field theory}

\vspace{6mm}

Daniel Kabat${}^1$ and Gilad Lifschytz${}^2$ 

\vspace{1mm}

${}^1${\small \sl Department of Physics} \\
{\small \sl Columbia University, New York, NY 10027} \\
{\small \tt  kabat@phys.columbia.edu}

\vspace{1mm}

${}^2${\small \sl Department of Mathematics and Physics} \\
{\small \sl University of Haifa at Oranim, Tivon 36006, Israel} 
{\small \tt giladl@research.haifa.ac.il}

\end{center}

\vskip 0.3 cm

\noindent
We propose that the entropy of de Sitter space can be identified with
the mutual entropy of a dual conformal field theory.  We argue that
unitary time evolution in de Sitter space restricts the total number
of excited degrees of freedom to be bounded by the de Sitter entropy,
and we give a CFT interpretation of this restriction.  We also clarify
issues arising from the fact that both de Sitter and anti de Sitter
have dual descriptions in terms of conformal field theory.
\end{titlepage}

\section{Introduction}

Thanks to AdS/CFT and M(atrix) theory, we have reasonable
non-perturbative formulations of quantum gravity in backgrounds with
negative or vanishing cosmological constant \cite{BFSS, Maldacena,
reviews}.  A natural next step is to study backgrounds with positive
cosmological constant.  However, despite considerable effort, our
understanding of de Sitter space is quite limited.  In part this is
due to the difficulty of finding de Sitter solutions to string theory.
But in part this is also due to the fact that a number of conceptual
issues remain to be understood.  Perhaps the foremost conceptual issue
is to understand the entropy of de Sitter space.

Our study of de Sitter space will be based on the dS/CFT
correspondence \cite{Strominger}; for other relevant work see
\cite{deSitter}.  We will propose an understanding of de Sitter
entropy in this context.  We are motivated by the idea that de Sitter
entropy can be understood as entropy of entanglement evaluated on
Cauchy surfaces in the bulk of de Sitter space.  In the dual CFT, we
will argue that de Sitter entropy can be understood as ``mutual
entropy'' -- a sort of Euclidean entropy of entanglement.  We will
also point out some connections between physics in de Sitter and AdS
spaces.  It is natural to expect some connections, since the de Sitter
metric can be obtained by an analytic continuation from AdS.  Indeed
this continuation may help to explain why both de Sitter and AdS are
dual to CFT's.

An outline of this paper is as follows.  In section 2 we review some
basic properties of de Sitter space and the dS/CFT correspondence.  In
section 3 we present our interpretation of de Sitter entropy in terms
of mutual entropy in the CFT.  In section 4 we discuss the analytic
continuation from AdS to de Sitter space, and use this to gain insight
into the dS/CFT correspondence.  In section 5 we present an alternate
interpretation of the entropy, based on the continuation from AdS,
which is appropriate to static coordinates.  In section 6 we discuss
the entropy bounds arising from unitary time evolution.

\section{de Sitter / CFT duality}

We begin by recalling some properties of de Sitter space.
$(d+1)$-dimensional de Sitter space $dS_{d+1}$ is the hyperboloid
\be
\label{dShyperboloid}
-(X^0)^2 + (X^1)^2 + \cdots + (X^{d+1})^2 = \ell^2
\ee
inside ${\mathbb R}^{d+1,1}$, where $\ell$ is the de Sitter radius.
$dS_{d+1}$ inherits its isometry group $SO(d+1,1)$ from this
embedding.  This group is also the conformal group in $d$ Euclidean
dimensions, which motivates the dS/CFT conjecture: the observables of
quantum gravity in de Sitter space can be obtained from a Euclidean
conformal field theory in one less dimension.  The CFT may well be
non-unitary \cite{Strominger}, although as we shall see it seems to
have many of the properties of a unitary CFT.

\begin{figure}
\epsfig{file=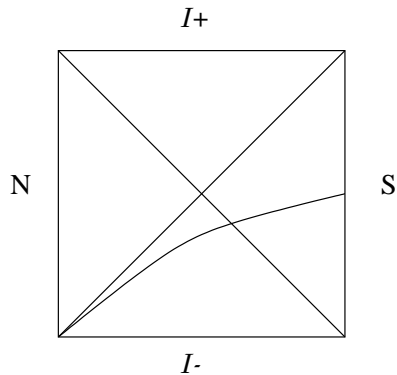}
\caption{Penrose diagram for de Sitter space.  We've indicated the two
cosmological horizons as well as a fixed-time hypersurface in the
planar coordinate system.}
\end{figure}

The Penrose diagram for de Sitter space is shown in Fig.~1.  A
comoving observer can only interact with a finite region of the spacetime.
For an observer at the south pole, this region is the right triangle
in Fig.~1.  This triangle can be described using the static metric
\be
\label{StaticMetric}
ds^2 = - \left(1 - {r^2 \over \ell^2}\right)dt^2 + \left(1 - {r^2 \over \ell^2}\right)^{-1}dr^2
+ r^2 d\Omega_{d-1}^2\,.
\ee
The observer is located at $r = 0$, while the horizon is the sphere at
$r = \ell$.  The horizon has a temperature $T = 1/2\pi\ell$, while the
entropy is given by the standard expression \cite{GibbonsHawking}
\be
\label{CosmologicalEntropy}
S = {{\rm area} \over 4G} = {{\rm vol}(S^{d-1}) \,\, \ell^{d-1} \over 4 G} \,.
\ee

We will also use planar coordinates, which cover the bottom and right
triangles in Fig.~1.  The metric is
\be
\label{PlanarMetric}
ds^2 = {\ell^2 \over \eta^2}\left(-d\eta^2 + dx^i dx^i\right)\,.
\ee
These coordinates parameterize the de Sitter hyperboloid (\ref{dShyperboloid}) by setting
\bea
\label{coords}
X^0 & = & {\eta \over 2} - {\ell^2 \over 2 \eta} - {1 \over 2 \eta} x^i x^i \\
\nonumber
X^i & = & \ell x^i / \eta \\
\nonumber
X^{d+1} & = & {\eta \over 2} + {\ell^2 \over 2 \eta} - {1 \over 2 \eta} x^i x^i \,.
\eea
The observer is located at $x^i = 0$, while the past horizon is located at $X^0 + X^{d+1} =
0$, or equivalently at
\be
\label{radius}
x^i x^i = \eta^2\,.
\ee
That is, in the planar coordinate system, the cosmological horizon is
a sphere of radius $\eta$.

\begin{figure}
\epsfig{file=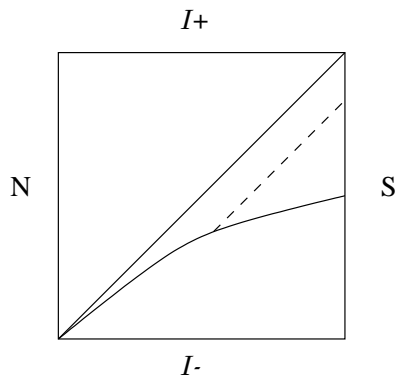}
\caption{The dashed line is a contracting light-sheet, which begins at
the south pole and intersects the constant-$\eta$ hypersurface at
$x^i x^i = R^2$.}
\end{figure}

We now discuss entropy bounds in planar coordinates.  At time $\eta$,
consider a spatial ball of arbitrary radius $x^i x^i \leq R^2$.
Following Bousso's formulation of the holographic principle
\cite{Bousso}, the entropy contained in this ball is bounded by the
entropy on the contracting light-sheet shown in Fig.~2, which in turn
is bounded by
\be
\label{Entropy}
S = {\hbox{\rm surface area of ball} \over 4 G}
= {{\rm vol}(S^{d-1}) \over 4 G} \left({\ell R \over \eta}\right)^{d-1}\,.
\ee
When $R = \eta$, this reproduces the entropy of de Sitter space
(\ref{CosmologicalEntropy}).

Planar coordinates are particularly convenient for discussing
dS/CFT duality \cite{Strominger}.  We take the continuum Euclidean CFT
to live at past infinity ($\eta = 0$), on the space parameterized by
$x^i$.  We identify the de Sitter isometry group with the group of
conformal transformations of the CFT.  For example, the de Sitter
metric is invariant under
\be
\label{isometry}
\eta \rightarrow \lambda \eta \qquad\quad x^i \rightarrow \lambda x^i\,.
\ee
This isometry of de Sitter space corresponds to a dilation of the CFT,
$x^i \rightarrow \lambda x^i$.  The de Sitter invariant vacuum state
of the gravity theory should correspond to the $SO(d+1,1)$-invariant
ground state of the CFT.

The de Sitter time coordinate $\eta$ arises holographically, and is
not manifest in the CFT.  To understand its significance, we use the
UV/IR correspondence.  This states that an object at time $\eta$ in de
Sitter space corresponds to an excitation of scale size
\be
\label{UVIR}
\vert \delta x \vert = \eta
\ee
in the CFT.  This is the de Sitter analog of the UV/IR
relation familiar in AdS/CFT duality \cite{SusskindWitten,PeetPolchinski}.

Since objects on a constant-time hypersurface have a finite scale size
in the CFT, we can encode them in a CFT with a short-distance cutoff.
We will discuss the cutoff procedure in more detail in section 3, but
the most straightforward procedure is to imagine that the CFT is put
on a Euclidean lattice with lattice spacing $a$.  This regulated CFT
contains all the degrees of freedom necessary to describe the
hypersurface $\eta = a$ in de Sitter space.

\section{de Sitter entropy in planar coordinates}

From the point of view of bulk quantum gravity, it seems natural to
associate an entropy with de Sitter space.  This can be seen as
follows.  In the planar coordinate system (\ref{PlanarMetric}),
consider a fixed-time hypersurface $\eta = {\rm const}$.  This is a
Cauchy surface.  But an observer located at the south pole can only
interact with those degrees of freedom which are located on a part of
this hypersurface, the part inside the cosmological horizon.  Even if
empty de Sitter space is associated with a pure state on this Cauchy
surface, physics inside the horizon would have to be described in
terms of a density matrix.  It seems natural to identify the entropy
of de Sitter space (\ref{CosmologicalEntropy}) with the entropy of
this density matrix.  That is, from the point of view of bulk quantum
gravity, we would like to understand de Sitter entropy as arising from
entropy of entanglement \cite{Bombelli, Srednicki}.

More generally, the holographic bound (\ref{Entropy}) allows us to
associate an entropy with a ball of arbitrary radius $x^i x^i \leq
R^2$ on the hypersurface $\eta = {\rm const}$.  This generalization
will turn out to be very useful for us.  The two entropies
(\ref{CosmologicalEntropy}) and (\ref{Entropy}) are quite distinct and
should not be confused.  The cosmological entropy
(\ref{CosmologicalEntropy}) is an intrinsic property of de Sitter
space, arising from its global causal structure, and limits the number
of degrees of freedom that any observer can interact with.  The
holographic bound (\ref{Entropy}), on the other hand, refers (for $R >
\ell$) to degrees of freedom that are outside the cosmological
horizon.  An observer at the south pole will eventually be able to see
these degrees of freedom, but can never influence them (see Fig.~2).
The two notions of entropy coincide if one sets $R = \eta$, so that
the surface of the ball coincides with the cosmological horizon.

We will assume that the holographic bound (\ref{Entropy}) is
meaningful, and ask what it corresponds to in the CFT.  The CFT lives
on a space parameterized by $x^i$.  Physics on the hypersurface $\eta
= {\rm const.}$ is described by a CFT with a short-distance cutoff
$\delta x = \eta$.  We will discuss the cutoff in more detail below,
but for now imagine putting the CFT on a Euclidean lattice with lattice
spacing equal to $\eta$.  We also assume that the CFT has a Lagrangian
description, so that it can be understood in terms of degrees of
freedom $\phi_{i,n}$.  Here $i$ labels the various lattice sites, and
$n$ labels the degrees of freedom at each site.  These degrees of
freedom will fluctuate according to a probability distribution
$P(\phi) = {1 \over Z} e^{-S(\phi)}$, where $S$ is the Euclidean
action.\footnote{It is not clear whether this is the correct
framework, since the CFT which is dual to de Sitter space might be
non-unitary.}

We are interested in the region $x^i x^i \leq R^2$ of the hypersurface
$\eta = {\rm const}$.  It seems natural to expect that this region of
the bulk spacetime is holographically encoded in a finite region of
the cutoff CFT, namely a ball $x^i x^i \leq R^2$.  Note that the
region inside the cosmological horizon should be encoded in a single
lattice site, see (\ref{radius}).

Our proposal is that the entropy (\ref{Entropy}) associated with
regions of de Sitter space can be understood as arising from
correlations between CFT degrees of freedom located on opposite sides
of the sphere $x^i x^i = R^2$.  These correlations can be
quantified in terms of ``mutual entropy,'' a concept which we will
define more precisely momentarily.  Mutual entropy can be thought of
as a Euclidean analog of entropy of entanglement.  Stated more
precisely, our proposal is that entropy of entanglement in the bulk of
de Sitter space is dual (in the sense of the dS/CFT correspondence) to
mutual entropy in the CFT.

This gives rise to a rather attractive picture of time evolution.
From the bulk de Sitter point of view, as time evolves new degrees of
freedom can flow in through the past cosmological horizon.
Nonetheless the total entropy inside the horizon remains constant.
From the CFT point of view the region inside the horizon corresponds
to a single lattice site, and time evolution corresponds to a
block-spin transformation.  As time (RG parameter) evolves, the single
lattice site corresponds to a larger and larger region of the CFT
spacetime.  Nonetheless the total entropy associated with the site
remains constant.  We also get an amusing picture of what an observer
means in the Euclidean CFT: an inertial observer in de Sitter space
sits on a single lattice site of the CFT.

\subsection{Mutual entropy}

Consider a collection of random variables, which we separate into two
sets of degrees of freedom $X$ and $Y$.  We will have in mind that we can measure
$X$ but are unable to observe $Y$.  From the joint probability
distribution $P(X,Y)$ we can construct probability distributions for
$X$ and $Y$.
\beas
P(X) & = & \int dY \, P(X,Y) \\
P(Y) & = & \int dX \, P(X,Y)
\eeas
Even if we could observe both $X$ and $Y$ we would have some lack of
{\em a priori} information due to the fact that $X$ and $Y$ fluctuate.
This lack of information can be measured by the entropy\footnote{If a
random variable takes on discrete values, one can define entropy by $H
= - \sum_k P_k \log P_k$.  In the continuum limit $P_k \rightarrow
P(x) dx$ and the entropy diverges.  As usual we define the entropy of
a continuous distribution by discarding this divergence.}
\[
H_{XY} = - \int dX dY \, P(X,Y) \log P(X,Y)\,.
\]
Likewise we can associate an entropy
\[
H_X = - \int dX \, P(X) \log P(X)
\]
with the fluctuations in $X$, and an entropy
\[
H_Y = - \int dY \, P(Y) \log P(Y)
\]
with the fluctuations in $Y$.  Now suppose that we can only
measure $X$.  Restricting to these degrees of freedom leads to an
increase in the entropy, due to the fact that we are not able to
observe correlations between $X$ and $Y$.  This increase can be
quantified by defining the mutual entropy (also known as mutual
information)
\be
\label{MutualEntropy}
I_{XY} = H_X + H_Y - H_{XY}\,.
\ee
Note that mutual information, since it involves entropy differences,
is free from the divergences mentioned in${}^2$.

Mutual entropy is a familiar concept in information theory
\cite{info}.  It satisfies a number of properties,
\begin{itemize}
\item
$I_{XY} \geq 0$, with equality iff $X$ and $Y$ are statistically
independent (that is, iff $P(X,Y) = P(X) P(Y)$).
\item
For discrete random variables
\be
\label{EntropyInequality}
I_{XY} \leq {\rm min}(H_X,H_Y)
\ee
with equality iff $X$ and $Y$ are perfectly correlated (that is, iff the
value of $X$ uniquely determines the value of $Y$, and visa versa).
\end{itemize}

\subsection{Mutual entropy in quantum field theory}

In this section we study mutual entropy in the context of Euclidean
quantum field theory.  Consider a quantum field which evolves from an
initial field eigenstate $\ket{q_i}$ at time $\tau_i$ to a final eigenstate
$\ket{q_f}$ at time $\tau_f$.  This is described by the configuration space
path integral
\[
Z = \langle q_f,\tau_f \vert q_i,\tau_i \rangle
= \int\limits_{\scriptstyle \phi(\tau_i) = q_i \atop \phi(\tau_f) = q_f} \!\!\! {\cal D} \phi \, e^{-S} \,.
\]
Cutting open the path integral at some intermediate time $\tau$, one has
\[
Z = \int dq \, \langle q_f,\tau_f \vert q,\tau \rangle
\langle q,\tau \vert q_i,\tau_i\rangle
\]
so the probability distribution for $q$ is
\[
P(q) = {1 \over Z} \langle q_f,\tau_f \vert q,\tau \rangle
\langle q,\tau \vert q_i,\tau_i\rangle\,.
\]
Now let us separate the field into
\beas
X & = & \lbrace\hbox{\rm degrees of freedom between $\tau_i$ and $\tau$}\rbrace \\
Y & = & \lbrace\hbox{\rm degrees of freedom between $\tau$ and $\tau_f$}\rbrace\,.
\eeas
In a continuum field theory it doesn't matter whether we regard $q$ as
belonging to $X$ or $Y$.  The joint probability distribution for these
degrees of freedom is
\[
P(X,Y) = {1 \over Z} e^{-S(X,Y)}
\]
with an associated entropy
\[
H_{XY} = - \int {\cal D}X{\cal D}Y \, P(X,Y) \log P(X,Y)
= \log Z + \langle S(X,Y) \rangle\,.
\]
The probability distribution for $X$ is
\[
P(X) = \int {\cal D}Y \, P(X,Y) = {1 \over Z} e^{-S(X)}
\langle q,\tau \vert q_i,\tau_i \rangle,
\]
where we have regarded $q$ as belonging to the set $X$, and have made
use of the locality axiom $S(X,Y) = S(X) + S(Y)$.  The entropy
associated with $X$ is then
\beas
H_X & = & - \int {\cal D}X \, P(X) \log P(X) \\
& = & - \int dq \, P(q) \log \left(
{1 \over Z} \langle q,\tau \vert q_i,\tau_i \rangle \right)
+ \int {\cal D}X{\cal D}Y \, P(X,Y) S(X)\,.
\eeas
Likewise the entropy for $Y$ is
\[
H_Y = - \int dq \, P(q) \log \left(
{1 \over Z} \langle q_f,\tau_f \vert q,\tau \rangle \right)
+ \int {\cal D}X{\cal D}Y \, P(X,Y) S(Y)\,,
\]
where now we're regarding $q$ as belonging to the set $Y$.  Putting these results
together, the mutual entropy is given by
\be
\label{MutualResult}
I_{XY} = - \int dq \, P(q) \log P(q)\,.
\ee
In this derivation we have implicitly used a configuration-space path
integral, so we take the probability distribution appearing in
(\ref{MutualResult}) to be determined by the wavefunction
of the system written in configuration space.

We will be most interested in the limit $\tau_i \rightarrow - \infty$
and $\tau_f \rightarrow +\infty$.  Then $P(q) \rightarrow
\psi_0^*\psi_0(q)$, where $\psi_0$ is the ground state wavefunction.
That is, in this limit mutual entropy is associated with the ground
state wavefunction of a system, and is obtained by regarding the
wavefunction as providing a statistical distribution $P(q)$ for the
possible configurations of the system at time $\tau$.\footnote{Mutual
entropy should not be confused with the usual notion of entropy in
quantum statistical mechanics, which vanishes in this case since
$\psi_0$ is a pure state.}

We are now in a position to compute the mutual entropy for a scalar
field of mass $m$.  For simplicity we work in two spacetime
dimensions, on a cylinder ${\mathbb R} \times S^1$.  Decomposing the
field into Fourier modes around the $S^1$ yields an infinite
collection of harmonic oscillators with frequencies $\omega_n =
\sqrt{\left({n / R}\right)^2 + m^2}$, where $n \in {\mathbb Z}$ and
$R$ is the radius of the spatial circle.  The entropy associated with
a harmonic oscillator in its ground state is $-{1 \over 2}\log
\omega$, so
\[
I_{XY} = - {1 \over 2} \sum_{n \in {\mathbb Z}} \log
\sqrt{\left({n \over R}\right)^2 + m^2}\,.
\]
The sum is divergent.  Subtracting the contribution of a Pauli-Villars regulator field
with mass $M \rightarrow \infty$ we have
\be
\label{ScalarEntropy}
I_{XY} = - {1 \over 4} \sum_{n \in {\mathbb Z}} \log {n^2 + m^2 R^2 \over
n^2 + M^2 R^2} = - {1 \over 2} \log {\sinh \pi m R \over \sinh \pi M R}\,.
\ee

\subsection{Mutual entropy in conformal field theory}

We now turn to the study of mutual entropy in conformal field theory,
and present a heuristic argument that mutual entropy is proportional
to the central charge of the CFT.

First let us count the degrees of freedom necessary to describe a
cut-off CFT.  One way to count the degrees of freedom with a given
cutoff $\Lambda$ is to study the system at a finite temperature $T
\sim \Lambda$.  The resulting thermal entropy counts the degrees of
freedom with energies up to the cutoff.  A two-dimensional unitary
CFT at temperature $T$ has an entropy per unit length given by
\cite{Cardy}
\be
{S \over L} = {\pi c T \over 3}
\ee
where $c$ is the central charge.  Since $L \Lambda$ is the number of spatial
lattice sites we see that the number of degrees of freedom per
lattice site needed to describe the cutoff theory is
proportional to the central charge.
\begin{equation}
\frac{S}{L\Lambda} \sim {\pi c \over 3}
\end{equation}

One can of course get the same result using the entropy-energy
relationship.  Let $x$ be the number of degrees of freedom per
lattice site.  Then one has
\be
\frac{S}{L\Lambda}\sim x,\ \ \frac{E}{L\Lambda} \sim x\Lambda
\ee
as well as the Cardy formula
\be
S \sim \sqrt{cEL}\,.
\ee
Again one finds $x \sim c$.

Thus the conformal field theory can be described in terms of
approximately $c$ degrees of freedom per lattice site.  We expect the
mutual entropy to be bounded by the available number of degrees of
freedom.  If the bound is saturated, we expect the mutual entropy to
be proportional to $c$.  In particular the entropy associated with a
single site in the lattice should be of order $c$.

As a concrete example of a conformally invariant theory,
consider a massless scalar field.  Using the result (\ref{ScalarEntropy}),
the entropy in the massless limit is given by
\[
I_{XY} = {\pi M R \over 2} - {1 \over 2} \log (2 \pi R M)
\]
where we have suppressed the contribution of the zero mode.  In radial
quantization, this is the mutual entropy of a $c=1$ CFT associated
with a disc of radius $R$, regulated with a Pauli-Villars cutoff.

\subsection{Mutual entropy and de Sitter entropy}

We now discuss the extent to which mutual entropy in a CFT has the
right properties to account for the entropy of de Sitter space.

For simplicity we will concentrate on the case of $dS_3/CFT_2$.  In this case the
central charge is known \cite{Strominger},
\[
c = {3 \ell \over 2 G}\,.
\]
Thus the holographic entropy (\ref{Entropy}) can be written as
\be
\label{HoloEntropy}
S = {\pi c R \over 3 \eta}\,.
\ee
The salient features are that the entropy is proportional to the central charge,
scales linearly with the radius $R$, and scales inversely with the time $\eta$.

As we discussed in section 3.3, it is reasonable to expect that the
mutual entropy of a conformal field theory is proportional to the
central charge.  Moreover, as we saw in section 3.2, the mutual
entropy of a two-dimensional field theory is linearly divergent.  For
example, for a scalar field with a Pauli-Villars cutoff the entropy is
\[
I_{XY} = {\pi M R \over 2}
\]
when $M R \gg 1$.  This displays the desired linear scaling with $R$.
The de Sitter CFT should have a short-distance cutoff at
$\delta x = \eta$, or equivalently at $M \approx 1 / \eta$.  So the
mutual entropy also has the desired inverse scaling with $\eta$.

Having mentioned these features, let us also point out some potential
difficulties.  One issue is that it is not clear whether the arguments
of section 3.3, that the mutual entropy of a CFT is proportional to
$c$, apply to non-unitary CFT's.  Another issue is that to obtain the
entropy associated with the cosmological horizon we need to set $R =
\eta$.  That is, we need to study the entropy associated with a region
in the CFT whose size is set by the UV cutoff.  The resulting
entropy will be non-universal, since it depends on exactly how one
regulates the CFT.  A related issue arises because the entropy is
linearly divergent.  The coefficient of a linear divergence is
non-universal, so it would seem that one could not hope to get the right
coefficient (the $\pi/3$ in (\ref{HoloEntropy})).

Let us mention how we think some of these difficulties are resolved.
We believe the de Sitter / CFT correspondence picks out a preferred
regulator for the CFT.  Roughly speaking, a slice $\eta = {\rm
const.}$ in de Sitter space can be described by a CFT with a short
distance lattice cutoff at $\delta x = \eta$.  However the precise
cutoff procedure should be given by using a bulk-to-boundary
propagator to map localized objects in the bulk to smeared-out
excitations on the boundary.  The size of the boundary excitations
will be roughly $\delta x = \eta$.  But the bulk-to-boundary
propagator should give a precise smearing function, or equivalently a
precise way of regulating the CFT.  If one could calculate mutual
entropy using this preferred regulator, one might be able to reproduce
the correct coefficient in the entropy.

A similar issue of regulator dependence arises in AdS/CFT duality.  In
the AdS/CFT context holography states that the number of degrees of
freedom of a cut-off CFT is equal to one quarter of the area of an
appropriate surface in Planck units \cite{SusskindWitten}.  To get the
one quarter a specific regularization scheme must be employed, but the
precise scheme has not been explicitly worked out.

\section{From AdS to de Sitter space}

Since the metric of AdS can be analytically continued to give the
metric of de Sitter space, and since the relationship between AdS and
CFT is well-understood, one might hope to use AdS/CFT duality to learn
more about dS/CFT.  In this section we adopt this approach, and use it
to gain some insight into the non-unitarity of the de Sitter CFT, as
well as the origins of supersymmetry breaking in de Sitter space.

First let us explain what we mean by analytically continuing from AdS
to de Sitter space.  The duality between string theory in an AdS
background and conformal field theory identifies the Hilbert space of
the gravity theory with the Hilbert space of the CFT \cite{BKLT}.
This identification implies the existence of a set of ``bulk
operators'' in the CFT, which we denote ${\cal{O}}_{\rm
AdS}(\tilde{t},\tilde{x},\tilde{r})$, where $(\tilde{t},\tilde{x})$
are the coordinates of the CFT and $\tilde{r}$ is the extra radial
coordinate of the AdS gravity theory.  These operators create states
in the CFT which are dual to localized excitations in the bulk.  As
long as we are in a regime where supergravity is valid, these bulk
operators suffice to describe quasi-local physics in the bulk of
AdS.\footnote{For a possible set of such bulk operators see
\cite{lifper}.}  We do not know how to construct a complete set of
bulk operators, but all we will use here is the fact that they exist.

These bulk operators depend on an insertion point in the bulk of AdS,
which is labeled by the coordinates of the CFT
$\tilde{t},\,\tilde{x}$ plus the extra radial coordinate
$\tilde{r}$. The bulk operators can be expanded in terms of CFT
operators, with coefficients depending on the CFT coordinates as well
as on the radial direction.
\begin{equation}
{\cal{O}}_{\rm AdS}(\tilde{t},\tilde{x},\tilde{r}) = \sum_{i} \int dtdx\,
f_{i}(t,x,\tilde{t},\tilde{x},\tilde{r}) O_{i}(t,x)
\end{equation}
Here the $O_{i}$ are a basis of local operators in the CFT.  Once one
has these operators one can compute their correlation functions, which
in principle encode all information about the bulk.  In particular
given the correlation functions one ought to be able to recover the
bulk metric.  If we start with a Lorentzian CFT on ${\mathbb R} \times
S^{d-1}$, then we would obtain the metric of AdS${}_{d+1}$, for
instance in the form (with signature $-+++\cdots$)
\begin{equation}
ds^2=-\left(1+\frac{\tilde{r}^2}{\ell^2}\right)d\tilde{t}^2 + \left(1+\frac{\tilde{r}^2}{\ell^2}\right)^{-1} 
d\tilde{r}^2 + \tilde{r}^2 d\Omega_{d-1}^{2}\,.
\label{adsm}
\end{equation}
These coordinates cover all of global AdS provided $-\infty< \tilde{t}
< \infty$. The AdS radius $\ell$ corresponds to a parameter in the
CFT.

Suppose we take the bulk operators
${\cal{O}}_{\rm AdS}(\tilde{t},\tilde{x},\tilde{r})$ and analytically continue them
to complex values of $\tilde{t}$ and $\tilde{r}$.  This allows us to define
\begin{equation}
{\cal{O}}_{\rm dS}(t,\tilde{x},r) \equiv {\cal{O}}_{\rm AdS}(it,\tilde{x},ir)\,.
\end{equation}
These double Wick rotated operators ${\cal{O}}_{\rm dS}(t,\tilde{x},r)$
are still formally operators in the CFT (now a Euclidean CFT), so in
principle one can compute their correlation functions and recover the
metric of the bulk spacetime.  But the metric that one will deduce from this
procedure is now
\begin{equation}
ds^2=\left(1-\frac{r^2}{\ell^2}\right)dt^2 - \left(1-\frac{r^2}{\ell^2}\right)^{-1} dr^2 -
r^2 d\Omega_{d-1}^{2}\,.
\label{dsm}
\end{equation}
This is the metric of Lorentzian de Sitter space, obtained in the
static coordinates (\ref{StaticMetric}), but with a flipped signature
($+---\cdots$).  The metric has a coordinate singularity at $r =
\ell$.  But by extending the range of the radial coordinate to $0 < r
< \infty$, we will regard these coordinates as covering one half of de
Sitter space, which includes either past infinity or future infinity.
The dual CFT then lives on the boundary $r \rightarrow \infty$, with
topology ${\mathbb R} \times S^{d-1}$.

Similar analytic continuations can be set up in other coordinate
systems.  For example, one can start from Euclidean AdS with metric
\begin{equation}
\label{EAdS}
ds^2=\frac{\ell^2}{z^2}(dz^2 + d\tau^2 + dx^{i}dx^{i})\,.
\end{equation}
Setting $\eta=iz$ gives
\begin{equation}
ds^2=\frac{\ell^2}{\eta^2}(d\eta^2 - d\tau^2 - dx^{i}dx^{i})
\end{equation}
which is just de Sitter space in planar coordinates with signature
$(+---\cdots)$.  These coordinates also cover half of de Sitter space, and
the corresponding CFT lives on ${\mathbb R}^{d}$.

Let us make some comments on these analytic continuations.  First of
all, even if one starts from a unitary CFT, it is quite possible that
the double Wick rotated operators are not in unitary representations
of the conformal group (for instance they may create non-normalizable
states).  Also note that we are ignoring the fate of the internal
compactification manifold, such as the $S^5$ of $AdS_5 \times S^5$.
The internal manifold can perhaps be dealt with along the lines of
\cite{Hull}.

In both examples we find that a Euclidean CFT naturally describes half
of de Sitter space.  The boundary of AdS is mapped to the asymptotic
past (or future) of de Sitter ($r \rightarrow \infty$ or $\eta
\rightarrow 0$), so one can regard the CFT as living at past (or
future) infinity.  The observables of the gravity theory live on the
boundary, and the description of the bulk of de Sitter is holographic.

To understand the continuation in more detail, consider a scalar field
of mass $m^2$ in Lorentzian anti de Sitter space, with equation of
motion
\begin{equation}
(\square_{\rm AdS} -m^2) \phi_{\rm AdS}=0 \,.
\label{sds}
\end{equation}
If we rotate $t = i\tilde{t}$, $r = i\tilde{r}$ and call the double
Wick rotated field $\phi_{\rm dS}$, it satisfies
\begin{equation}
(\square_{\rm dS} -m^2)\phi_{\rm dS}=0
\end{equation}
but now with a flipped signature, which gives the scalar field a mass
equal to $-m^2$.  This suggests that a scalar field in de Sitter space
with mass $+m^2$ corresponds to a scalar field in AdS with mass
$-m^2$.  This nicely explains the conformal dimensions of operators
that one gets from dS/CFT duality.  An operator dual to a field of
mass $m^2$ in de Sitter space has a conformal dimension
\cite{Strominger}
\begin{equation}
h_\pm = \frac{1}{2}(d\pm\sqrt{d^2 -4 m^2 \ell^2})\,.
\end{equation}
But this is the conformal dimension of an operator associated with a field
of mass $-m^2$ in AdS.

The full correspondence between fields in de Sitter and AdS, however, involves
more than just a double Wick rotation.  To see this, consider starting
in de Sitter space with a scalar field of mass $m^2$.  The Wightman
function for such a field is (in planar coordinates, in the so-called
Euclidean vacuum \cite{CandelasRaine})
\be
\label{dSWightman}
G_{\rm dS}(x,x') = F(h_+,h_-,{d + 1 \over 2},{1 + P(x,x') \over 2})
\ee
where
\beas
&& h_\pm = \frac{d}{2} \pm \sqrt{\biggl(\frac{d}{2}\biggr)^{2} - m^2 \ell^2} \\
&& P(x,x') = \frac{\eta^{2} + \eta'{}^2 - |x-x'|^2}{2 \eta \eta'}\,.
\eeas
When we analytically continue $z = -i \eta$ to obtain the Euclidean
AdS metric (\ref{EAdS}), $G_{\rm dS}$ does not turn into a Greens
function for a scalar field of mass $-m^2$ in AdS
\cite{BoussoMaloneyStrominger,SpradlinVolovich}.  Instead note that
the hypergeometric function satisfies
\begin{eqnarray}
\label{HyperId}
F(\alpha,\beta,\gamma,z)=(1-z)^{-\alpha}\frac{\Gamma(\gamma)\Gamma(\beta-\alpha)}{\Gamma(\beta)\Gamma(\gamma-\alpha)}F(\alpha,\gamma-\beta,\alpha-\beta+1,\frac{1}{1-z})\nonumber\\
+(1-z)^{-\beta}\frac{\Gamma(\gamma)\Gamma(\alpha-\beta)}{\Gamma(\alpha)\Gamma(\gamma-\beta)}F(\beta,\gamma-\alpha,\beta-\alpha+1,\frac{1}{1-z})\,.
\end{eqnarray}
Each term on the right hand side of this equation is proportional to
the Greens function for a scalar field in Euclidean AdS with mass
squared equal to $-m^2$ \cite{bcfm}.  That is, the de Sitter Greens
function becomes a sum of AdS Greens functions.

What does this imply for the relation between de Sitter space and AdS?
First consider a scalar field in de Sitter space with mass satisfying
\begin{equation}
\label{MassRange}
\frac{d^2}{4} - 1 < m^2_{\rm dS}\ell^2 < \frac{d^2}{4}\,.
\end{equation}
The mass of the corresponding field in AdS satisfies
\begin{equation}
\label{MassRange2}
- \frac{d^2}{4} < m^2_{\rm AdS}\ell^2 < 1 - \frac{d^2}{4}\,.
\end{equation}
In this mass range a scalar field in AdS can be quantized in two
inequivalent ways, corresponding to two possible choices of boundary
conditions at infinity \cite{bf,KlebanovWitten}.  The different
boundary conditions give rise to distinct Greens functions, and in
(\ref{HyperId}) {\it both} Greens functions appear.  This means that a
single field in de Sitter space must be associated with a pair of
fields in AdS.  In the dual AdS CFT this means we must have a pair of
operators, one of dimension $h_+$ and one of dimension
$h_-$.\footnote{From the point of view of the AdS CFT, the mass range
(\ref{MassRange2}) is the unitarity bound for these two operators:
$h_+,\,h_-$ real and larger than ${d \over 2} - 1$
\cite{KlebanovWitten}.}  Going back to de Sitter space, a single
scalar field in de Sitter space should correspond to a bulk operator
${\cal O}_{\rm dS}$ in the de Sitter CFT.  ${\cal O}_{\rm dS}$ can be
constructed by taking a linear combination of the two corresponding
bulk operators in the AdS CFT and performing a Wick rotation $z = -i
\eta$.

Outside the mass range (\ref{MassRange}) the situation is more subtle.
The identity (\ref{HyperId}) still holds, but there is only a
single quantization of a scalar field in AdS, and one or both of the
operators in the AdS CFT will violate the unitarity bound.

To summarize, each field in the mass range (\ref{MassRange}) in de
Sitter space corresponds to a pair of fields in AdS.  Having both
fields in AdS breaks supersymmetry, since AdS supersymmetry requires
particular boundary conditions on the fields \cite{bf,haw}.  The de
Sitter CFT bulk operators are linear combinations of the corresponding
Wick rotated AdS CFT bulk operators.

\section{Entropy in static coordinates}

In this section we discuss the entropy of de Sitter space from the
point of view of the CFT on ${\mathbb R} \times S^{d-1}$, {\em i.e.}
in the static $r,\,t$ coordinate system.  First let us recall how one
computes the entropy of de Sitter space.  One takes a smooth Euclidean
section of de Sitter space, which is just the round sphere.  In static
coordinates this is done by rotating $t$ to $it$.  To get a smooth
metric one has to periodically identify imaginary time with period
$2\pi\ell$.  One is left with a space parameterized by $0< it <
2\pi\ell$, $0 < r < \ell$, $\Omega \in S^{d-1}$.  Then one computes
the Euclidean action, which is minus the entropy (the energy is zero).

Analytically continuing this procedure to AdS it seems we are throwing
out the $\tilde{r}>\ell$ part of AdS.  The resulting entropy should be
interpreted as entropy of entanglement between the two regions
$\tilde{r}>\ell$ and $\tilde{r}<\ell$ in the AdS vacuum state.  In the
dual CFT the $\tilde{r}$ coordinate arises holographically, and is related to
the cutoff on the CFT
\cite{SusskindWitten}.  So from the CFT point of view this entropy
does not arise from entanglement between two regions of space, but
rather arises from entanglement between different energy
scales.

Entanglement entropy between different energy scales (unlike between
two regions of space) vanishes for a free theory, since a free theory
has a vacuum wavefunctional which is a tensor product in momentum
space.  But in an interacting theory the appearance of non-trivial
higher-point correlation functions implies entanglement between energy
scales.  In order to actually compute this entropy from the CFT we
would have to know the vacuum wavefunctional of the CFT.  Lacking
this, we use the fact that the CFT is equivalent to a semi-local field
theory in the bulk of AdS.  If the bulk theory was really local then
the entanglement entropy between $\tilde{r}>\ell$ and $\tilde{r}<\ell$
would be UV divergent, and proportional to the area of the boundary
between the two regions.  This reflects the fact that entanglement
entropy is measuring correlations between the two regions, given some
cutoff scale.  As the bulk theory is not really local one cannot take
the UV cutoff to zero.  A natural cutoff in AdS is provided by the
Planck length, so one expects the entropy of entanglement to be
proportional to
\begin{equation}
\frac{A}{G}=\frac{{\rm vol}(S^{d-1})\, \ell^{d-1}}{G}\,.
\end{equation}
Without further information we cannot deduce the coefficient, but we
can put a bound on it. The entanglement entropy between the two
regions cannot exceed the log of the number of degrees of freedom
residing in either one of the regions, since the entanglement entropy
can be computed from a density matrix in either region.  From AdS/CFT
we know that gravity in the region $\tilde{r}<\ell$ is described by a
finite number of degrees of freedom, while gravity in the region
$\tilde{r}>\ell$ has an infinite number of degrees of freedom.  This
gives an upper bound on the entanglement entropy, namely the number of
degrees of freedom residing at $\tilde{r}<l$, which is given by
\begin{equation}
\frac{{\rm vol}(S^{d-1})\, \ell^{d-1}}{4G}\,.
\end{equation}
This is exactly the de Sitter entropy.  Thus de Sitter entropy
saturates the bound, and the entanglement entropy is maximal.

The degrees of freedom within one AdS radius can be described by a
dimensionally reduced theory \cite{Flat}.  This is very similar to the
picture we discussed in section 3, where the region inside the horizon
of de Sitter space is described by a single lattice site.  It would be
very interesting to understand the relation between the two pictures
of the entropy we have developed, for planar coordinates (dual to CFT
on ${\mathbb R}^d$) and static coordinates (dual to CFT on ${\mathbb
R} \times S^{d-1}$).

\section{Entropy and time evolution}

We have seen that the entropy of de Sitter space can be understood as
entropy of entanglement between energy scales in the vacuum state of
the CFT on ${\mathbb R} \times S^{d-1}$, or in terms of mutual entropy
in the CFT on ${\mathbb R}^d$.  We now discuss a dynamical role which
the entropy plays in de Sitter space \cite{BanksBousso}.

Unitary time evolution requires that two different initial states will
evolve to two different final states.  However since time evolution in
de Sitter space is associated with changing the cutoff of the CFT, at
first sight it seems that unitary evolution is not possible.  To
understand what unitarity means let us consider the CFT on ${\mathbb
R} \times S^{d-1}$, which is dual to de Sitter space in static
coordinates.
\be
ds^2 = - \left(1 - {r^2 \over \ell^2}\right) dt^2 + \left(1 - {r^2 \over \ell^2}\right)^{-1}dr^2 +
r^2 d\Omega_{d-1}^2\,.
\ee
For $r > \ell$ surfaces of constant $r$ are space-like Cauchy
surfaces, and the Euclidean time $t$ of the CFT is a space-like
coordinate.  A surface of constant $r$ corresponds to a CFT with a
cutoff, for example a lattice with spacing determined by $r$.  As $r$
is decreased the lattice spacing gets bigger, and eventually the whole
$S^{d-1}$ is taken up by a single lattice site.  This happens when $r
= \ell$.  At this point the CFT effectively lives on a one dimensional
lattice in the $t$ direction, with lattice spacing equal to
$\ell$.\footnote{To enter the region $r < \ell$ one should keep an
appropriate subset of the degrees of freedom on these remaining lattice
sites.  This is a subtle problem, discussed for example in
\cite{Flat}.}

Now let's regard the CFT from a Hamiltonian point of view.  At any
Euclidean time $t$ we can determine the state of the CFT, for example
by doing a Euclidean path integral, possibly with some operator
insertions if the bulk de Sitter space is not in its vacuum state.
The de Sitter spacetime corresponds to a whole history of CFT states,
one state at each value of Euclidean time.  The question is what CFT
states are allowed to be part of this history.

de Sitter time evolution corresponds to RG transformations in the CFT.
We want to be able to coarsen the lattice to the point where $r =
\ell$, and still distinguish the different histories of the CFT.  But
as we saw in section 3.3, a CFT only has approximately $c$ degrees of
freedom per lattice site.  This puts a limit on the number of
histories that we can distinguish.  Note that $c$ is also proportional
to the de Sitter entropy.  Thus de Sitter entropy is a measure of how
many states in the CFT can be involved in a description of unitary
time evolution in de Sitter space.

These unitarity bounds also appear in the analytic continuation from de
Sitter to AdS that we discussed in section 4.  To see this, we first
recall some properties of the AdS/CFT correspondence.  Consider a
Lorentzian CFT on ${\mathbb R} \times S^{d-1}$.  This CFT is dual to
global AdS space.  AdS states with energy $E > M_{\rm bh}$, where
\begin{equation}
M_{\rm bh}=\frac{(d-1)\, {\rm vol}(S^{d-1}) \, \ell^{d-2}}{8\pi G}\,,
\end{equation}
will form a stable black hole, while states with energy $E < M_{\rm
bh}$ will at most form an unstable black hole that subsequently decays
\cite{hawkingpage,witten2}.  Stability is determined by comparing the
Euclidean action of the black hole to the free energy of a thermal
gas.  In the coordinates we are using, the black hole is stable if the
horizon radius $r_0$ satisfies $r_0 > \ell$.  Note that the smallest
stable AdS black hole (the one with $r_0 = \ell$) has an entropy ${\rm
vol}(S^{d-1}) \ell^{d-1} / 4G$ which is exactly equal to the entropy
of de Sitter space.  So if one gives the AdS CFT an entropy which
exceeds the entropy of the corresponding de Sitter space, a stable
black hole forms inside AdS.

What does this mean when we analytically continue to de Sitter space?
The metric of the AdS black hole is
\beas
& & ds^2 = - \left(1 + {\tilde{r}^2 \over \ell^2} - {w_d M \over \tilde{r}^{d-2}}\right) d\tilde{t}^2
+ \left(1 + {\tilde{r}^2 \over \ell^2} - {w_d M \over \tilde{r}^{d-2}}\right)^{-1} d\tilde{r}^2 + \tilde{r}^2 d\Omega_{d-1}^2 \\
& & w_{d} = \frac{16\pi G}{(d-1) \, {\rm vol(S^{d-1})}}
\eeas
and the substitution $r= i\tilde{r}$ gives three
possibilities.
\begin{itemize}
\item For odd $d$ one gets an imaginary mass for the de Sitter black hole.
\item For $d = 4n + 2$ one gets a black hole in de Sitter space with a
negative mass, which gives a naked singularity.
\item For $d=4n$ one gets a positive mass black hole in de Sitter
space.  However the mass of a de Sitter black hole is bounded above,
$M < M_{\rm max}$, by the requirement that the black hole horizon is smaller
than the cosmological horizon (otherwise a naked singularity appears).
One finds that $M_{\rm bh} > M_{\rm max}$, so even in this case one gets a
naked singularity.
\end{itemize}
Thus in all cases, a stable black hole in AdS space corresponds to a naked
singularity in de Sitter space.

This shows that if we exceed the mass (or entropy) limit in the CFT
naked singularities will form in de Sitter space.\footnote{This has
also been argued in \cite{singular}.}  These singularities can be
thought of as symptoms of the breakdown of unitary time evolution.  On
the CFT side exceeding the entropy bound means the CFT undergoes a
phase transition \cite{witten2}.  Thus only the low energy phase of
the CFT on ${\mathbb R} \times S^{d-1}$ describes de Sitter space.

\bigskip
\goodbreak
\centerline{\bf Acknowledgements}
We are grateful to Ofer Aharony, Micha Berkooz, Sumit Das, Nori
Iizuka, David Lowe, Don Marolf, Yaron Oz, Soo-Jong Rey, Cobi
Sonnenschein and Rafael Sorkin for valuable discussions.  This work is
supported by US-Israel Binational Science Foundation grant \#2000359.
DK is also supported in part by the DOE under contract
DE-FG02-92ER40699.

\end{document}